\documentclass{article}
\usepackage{epsfig}
\usepackage{amsmath}
\usepackage{amssymb}
\title{Dynamics of inelastically colliding spheres with Coulomb
  friction: Relaxation of translational and rotational energy}
 \author{Olaf Herbst, Martin Huthmann, and Annette
  Zippelius \\ {\em Institut f\"ur Theoretische Physik, Universit\"at
    G\"ottingen},\\{\em 
  Bunsenstr. 9, 37073 G\"ottingen, GERMANY}} 
\date{\today}
 
\begin{document}
 
\maketitle
\begin{abstract}
  We investigate the free cooling of inelastic rough spheres in the
  presence of Coulomb friction.  Depending on the coefficients of
  normal restitution $\epsilon$ and Coulomb friction $\mu$, we find
  qualitatively different asymptotic states.  For nearly complete
  normal restitution ($\epsilon$ close to 1) and large $\mu$, friction
  does not change the cooling properties qualitatively compared to a
  constant coefficient of tangential restitution.  In particular, the
  asymptotic state is characterized by a constant ratio of rotational
  and translational energies, both decaying according to Haff's law.
  However, for small $\epsilon$ and small $\mu$, the dissipation of
  rotational energy is suppressed, so that the asymptotic state is
  characterized by constant rotational energy while the translational
  energy continues to decay as predicted by Haff's law.  Introducing
  either surface roughness for grazing collisions or cohesion forces
  for collisions with vanishing normal load, causes the rotational
  energy to decay according to Haff´s law again in the asymptotic
  long-time limit with, however, an intermediate regime of
  approximately constant rotational energy.
 \end{abstract}

\section{Introduction}
Impact properties of small grains have been measured by several groups
\cite{Foerster,Experiments}.  The experimental data is frequently
parametrized using a simple model introduced by Walton \cite{Walton}.
The model involves three parameters; the first one, $\epsilon$
characterizes the incomplete restitution of the normal component of
the relative velocity of the contact point, denoted by
$\boldsymbol{g}$.  The second one, Coulomb's coefficient of friction
$\mu$ describes the reduction of the tangential component of
$\boldsymbol{g}$ due to sliding, while the third parameter, $\beta_0$
accounts for the incomplete restitution of the tangential component of
$\boldsymbol{g}$ for sticking contacts.  All three parameters have
been measured experimentally for various materials making a well
calibrated model available for theoretical investigations.

\section{Binary collisions}

We have previously investigated the free cooling of rough spheres
\cite{martin97,Luding} and needles \cite{martin99} using the formalism
of a Pseudo-Liouville operator. In this paper we extend the analysis
to include Coulomb friction.  We briefly recall the collision rules
for two spheres of equal diameter $a$, mass $m$ and moment of inertia
$I$. The unit-vector from the center of sphere two
($\boldsymbol{r}_{2}$) to the center of sphere one $($\mbox{\boldmath
  $r$}$_{1})$ is denoted by $\hat{\mbox{\boldmath$n$}} :=
(\boldsymbol{r}_{1}-\boldsymbol{r}_{2})/|\boldsymbol{r}_{1}
-\boldsymbol{r}_{2}|$. Center-of-mass velocities and angular
velocities before collision are denoted by $\mbox{\boldmath$v$}_1$,
$\mbox{\boldmath$v$}_2$, $\mbox{\boldmath$\omega$}_1$ and
$\mbox{\boldmath$\omega$}_2$.  Post-collisional quantities are primed.
The relative velocity of the contact point is given by
$\mbox{\boldmath$g$}=\mbox{\boldmath$v$}_1-\mbox{\boldmath$v$}_2
+\frac{a}{2}\hat{\mbox{\boldmath$n$}}\times(
\mbox{\boldmath$\omega$}_1 +\mbox{\boldmath$\omega$}_2)$.  The
relative velocity after collision is given by
 \begin{align}\label{rauku4}
   \hat{\mbox{\boldmath$n$}}\cdot\mbox{\boldmath$g$}^{'} =&
   -\epsilon(\boldsymbol{g},
   \hat{\boldsymbol{n}})(\hat{\mbox{\boldmath$n$}}\cdot\mbox{\boldmath$g$})\ 
   \quad {\rm with} \quad \epsilon(\boldsymbol{g},
   \hat{\boldsymbol{n}}) \in [0,1] , \\ \label{rauku5}
   \hat{\mbox{\boldmath$n$}}\times\mbox{\boldmath$g$}^{'} =&
   -\beta(\boldsymbol{g},
   \hat{\boldsymbol{n}})(\hat{\mbox{\boldmath$n$}}\times\mbox{\boldmath$g$})\ 
   \quad {\rm with} \quad \beta(\boldsymbol{g}, \hat{\boldsymbol{n}})
   \in [-1,1]
 \end{align}
 where $\epsilon(\boldsymbol{g}, \hat{\boldsymbol{n}})$ and
 $\beta(\boldsymbol{g}, \hat{\boldsymbol{n}})$ are the coefficients of
 restitution, which in general depend on $\boldsymbol{g}$ and $
 \hat{\boldsymbol{n}}$. We assume $\epsilon$ to be constant and allow
 $\beta$ to depend on the angle $\gamma$ between $\boldsymbol{g}$ and
 $ \hat{\boldsymbol{n}}$ in order to account for the different energy
 loss mechanisms of sliding and sticking contacts. The impact angle
 satisfies $\gamma \in [\frac{\pi}{2},\pi]$, so that
 $\cos{\gamma}=\hat{\boldsymbol{n}}\cdot
 \boldsymbol{g}/|\boldsymbol{g}|<0$.
 
 The two constitutive equations (\ref{rauku4},\ref{rauku5}) plus the
 conservation laws for linear and angular momenta determine the
 post-collisional velocities
\begin{align} \nonumber
  \mbox{\boldmath$v$}_1^{'} = \mbox{\boldmath$v$}_1+\boldsymbol{\Delta
    v},& \quad \nonumber \mbox{\boldmath$v$}_2^{'} =
  \mbox{\boldmath$v$}_2-\boldsymbol{\Delta v},\\ \label{freddy}
  \mbox{\boldmath$\omega$}_1^{'} =
  \mbox{\boldmath$\omega$}_1+\boldsymbol{\Delta \omega},& \quad
  \mbox{\boldmath$\omega$}_2^{'} =
  \mbox{\boldmath$\omega$}_2+\boldsymbol{\Delta \omega}
 \end{align}
 where
 \begin{align}
   \label{Deltas}\nonumber
   \boldsymbol{\Delta v} =& - \frac{(1+\epsilon)}{2}
   (\hat{\mbox{\boldmath$n$}}\cdot\mbox{\boldmath$v$}_{12})
   \hat{\mbox{\boldmath$n$}}-
   \eta\hat{\mbox{\boldmath$n$}}\times\left(\mbox{\boldmath$v$}_{12}\times\hat{\mbox{\boldmath$n$}}
     +\frac{a}{2}
     \mbox{\boldmath$\omega$}_{12}\right)\\
   \boldsymbol{\Delta \omega} =& \frac{2 \eta}{q
     a}\left(\hat{\mbox{\boldmath$n$}}\times\mbox{\boldmath$v$}_{12} +
     \frac{a}{2}\hat{\mbox{\boldmath$n$}}\times
     (\hat{\mbox{\boldmath$n$}}\times
     \mbox{\boldmath$\omega$}_{12})\right)
 \end{align}
 with
 $\mbox{\boldmath$v$}_{12}=\mbox{\boldmath$v$}_1-\mbox{\boldmath$v$}_2
 $, $\boldsymbol{\omega}_{12} =
 \boldsymbol{\omega}_{1}+\boldsymbol{\omega}_{2}$, $\eta =
 \eta(\gamma):= q(1+\beta(\gamma))/(2(1+q))$, and $ q:= (4I)/(m a^2)$
 ($q = 0.4$ for homogeneous spheres).
 
\begin{figure}[htbp]
  \begin{center}
    \leavevmode \epsfig{file=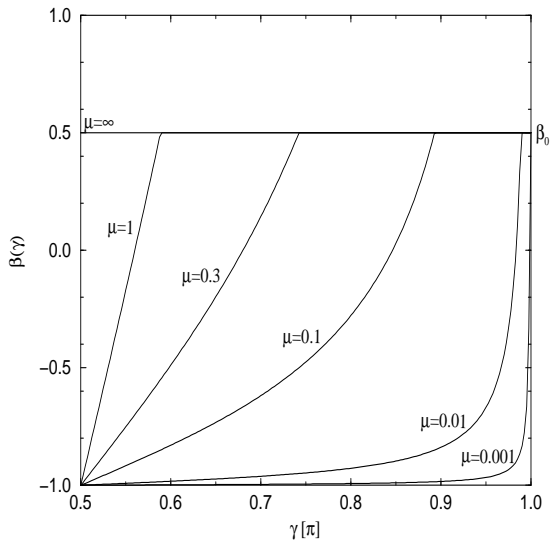,
      width=0.6\textwidth,height=0.6\textwidth}
    \caption{$\beta$($\gamma$) for different values of
      $\mu$; $\epsilon = 0.5$, $\beta_0 = 0.5$.}
    \label{BetaVonGamma}
  \end{center}
\end{figure}

 Sliding contacts are governed by Coulomb friction, giving rise to an
 impact angle dependent coefficient of tangential restitution
 \begin{equation}\label{betagamma}
   \beta^{*}(\gamma) = -1 - \frac{1+q}{q} (1+\epsilon) \mu
   \cot{\gamma}.
 \end{equation}
 Sliding, however, only occurs for impact angles $\gamma < \gamma_0$.
 For higher values one expects sticking which is governed by a
 constant coefficient of tangential restitution $-1<\beta_0\le1$ such
 that $\beta_0$ is strictly $>-1$.  In agreement with Walton
 \cite{Walton} we assume that either sliding or sticking occurs in any
 single collision, never both. We require $\beta (\gamma)$ to be
 continuous and obtain for the limiting angle
 $\gamma_0=\gamma_0(\epsilon, \beta_0, \mu)$
 \begin{equation}\label{gamma0x}
   \cot{\gamma_0} = - \frac{q}{1+q} \frac{1+\beta_0}{1+\epsilon}
   \frac{1}{\mu}
 \end{equation}
 so that
 \begin{equation}\label{betax}
   \beta(\gamma) = \min{\{\beta_0, -1 - \frac{1+q}{q} (1+\epsilon) \mu
     \cot{\gamma}\}}
 \end{equation}
 As shown in Fig.(\ref{BetaVonGamma}), $\mu \to 0$ corresponds to
 smooth spheres and $\mu \to \infty$ amounts to constant tangential
 restitution.

\section{Free Cooling of the many particle system}
\subsection{Analytical Theory}
We consider a system of $N$ classical particles confined to a
3-dimensional volume $V$ interacting through a hard-core potential.
The time evolution of a dynamic variable
$A=A(\{\mbox{\boldmath$r$}_k(t),\mbox{\boldmath$v$}_k(t),\mbox{\boldmath$\omega$}_k(t)\})$
is determined by a pseudo-Liouville operator ${\cal L}_{+}$ for $t >
0$
\begin{equation}
  A(\{\mbox{\boldmath$r$}_k,\mbox{\boldmath$v$}_k,\mbox{\boldmath$\omega$}_k\},t)
  =\exp(i{\cal
    L}_{+}t)A(\{\mbox{\boldmath$r$}_k,\mbox{\boldmath$v$}_k,\mbox{\boldmath$\omega$}_k\},0).
\end{equation}
 The pseudo-Liouville operator ${\cal L}_{+}$ consists of two parts
 ${\cal L}_{+}={\cal L}_0 +{\cal L}^{'}_{+}$. The first one, ${\cal
   L}_0$ describes the free streaming of particles
\begin{equation}
  {\cal L}_0 = -i \sum_k \mbox{\boldmath$v$}_k \cdot
  {\nabla}_{\mbox{\boldmath$\scriptstyle r$}_k},
\end{equation}
 and the second one, ${\cal L}^{'}_{+}=\frac{1}{2}\sum_{k\ne
   l}T_{+}(kl)$ describes hard-core collisions of two particles
\begin{equation} 
 \label{stossit}
 T_{+}(kl) =
 i(\mbox{\boldmath$v$}_{kl}\cdot\hat{\mbox{\boldmath$r$}}_{kl})
 \Theta(-\mbox{\boldmath$v$}_{kl}\cdot\hat{\mbox{\boldmath$r$}}_{kl} )
 \delta(|\mbox{\boldmath$r$}_{kl}|-a)(b_{kl}^{+}-1).
\end{equation}
The operator $b_{kl}^+$ replaces the linear and angular momenta of two
particles $k$ and $l$ before collision by the corresponding ones after
collision, according to eqs. (\ref{freddy}).  $\Theta(x)$ is the
Heaviside step--function, and we have introduced the notation
$\mbox{\boldmath$r$}_{kl}=\mbox{\boldmath$r$}_k-\mbox{\boldmath$r$}_l$
and $\hat{\mbox{\boldmath$r$}}_{kl}
=\mbox{\boldmath$r$}_{kl}/|\mbox{\boldmath$r$}_{kl}|$.  Equation
(\ref{stossit}) has a simple interpretation. The factor
$\mbox{\boldmath$v$}_{kl}\cdot\hat{\mbox{\boldmath$r$}}_{kl}$ gives
the flux of incoming particles. The $\Theta$- and $\delta$-functions
specify the conditions for a collision to take place. A collision
between particles $k$ and $l$ happens only if the two particles are
approaching each other which is ensured by
$\Theta(-\mbox{\boldmath$v$}_{kl}\cdot\hat{\mbox{\boldmath$r$}}_{kl}
)$. At the instant of a collision the distance between the two
particles has to vanish expressed by
$\delta(|\mbox{\boldmath$r$}_{kl}|-a)$. Finally, $(b_{kl}^{+}-1)$
generates the change of linear and angular momenta.

The ensemble average of a dynamic variable is defined by
 \begin{equation}\label{ensav1}
 \begin{split}
   \langle A \rangle _t &= \int d\Gamma \rho (0) A(t)=\int d\Gamma
   \rho
   (t) A(0)\\
   &= \int \prod_k(d\mbox{\boldmath$r$}_k d\mbox{\boldmath$v$}_k
   d\mbox{\boldmath$\omega$}_k) \rho (t) A(0).
 \end{split}
 \end{equation}
 Here $\rho(t)=\exp {(-i{\cal L}_+^{\dagger}t)}\,\rho(0)$ is the
 $N$-particle distribution function, whose time development is
 governed by the adjoint ${\cal L}_+^{\dagger}$ of the time evolution
 operator ${\cal L}_+$.  Differentiating equation (\ref{ensav1}) with
 respect to time we get
 \begin{equation}\label{ensav2}
 \begin{split}
   \frac{d}{dt} \langle A \rangle _t &= \int d\Gamma \rho (0)
   \frac{d}{dt}
   A(t)=\int d\Gamma \rho (0) i{\cal L}_+ A(t)\\
   &=\int d\Gamma \rho (0)
   \exp{(i {\cal L}_+ t)} i{\cal L}_+ A(0)\\
   &=\int d\Gamma \rho (t) i{\cal L}_+ A(0) = \langle i{\cal L}_+ A
   \rangle _t
 \end{split}
 \end{equation}
 We are interested in the average translational and rotational
 energies per particle
 \begin{align}
   E_{ {\rm tr}} =& \frac{1}{N}\sum_i\frac{m}{2}\mbox{\boldmath$v$}_i^2\\
   E_{ {\rm rot}} =&
   \frac{1}{N}\sum_i\frac{I}{2}\mbox{\boldmath$\omega$}_i^2
 \end{align}
 as well as the total kinetic energy $E=E_{ {\rm tr}}+E_{ {\rm rot}}$.
 
 We assume a {\it homogeneous cooling state} (HCS) and approximate the
 $N$-particle distribution function by a Gaussian
 \begin{equation}\label{hcs}
   \rho(t) \propto \prod_{k<l}\Theta(|\boldsymbol{r}_{kl}|-a)
   \exp{\left\{-\left(\frac{E_{{\rm tr}}}{T_{{\rm tr}}(t)}
+\frac{E_{{\rm rot}}}{T_{{\rm rot}}(t)}\right)\right\}}
 \end{equation}
 where the product of Heaviside functions accounts for the excluded
 volume.  The state of the system depends on time only through the
 average translational and rotational kinetic energies.  Hence its
 full time dependence (within the HCS approximation) is determined by
 two coupled differential equations for $T_{{\rm tr}}(t)$ and $T_{{\rm
     rot}}(t)$
\begin{align}
  \frac{3}{2} \frac{d}{dt} T_{{\rm tr}}(t) =& \frac{d}{dt}\langle E_{
    {\rm tr}}\rangle_{t} = \langle i{\cal L}_{+}E_{ {\rm tr}}\rangle_{t}\\
  \frac{3}{2} \frac{d}{dt} T_{{\rm rot}}(t) =& \frac{d}{dt}\langle E_{
    {\rm rot}}\rangle_{t} = \langle i{\cal L}_{+}E_{ {\rm
      rot}}\rangle_{t}
 \end{align}
 The expectation values on the right hand side can be calculated for
 the HCS state.  We obtain
 \begin{multline} 
 \label{dgl1}
 \frac{1}{\nu} \frac{d}{dt} T_{{\rm tr}}(t) = - T^{3/2}_{{\rm
     tr}}\left\{
   \frac{1-\epsilon^2}{4} \right. \\
 - \left. \eta^2_0 \frac{(1+\frac{T_{{\rm rot}}}{q T_{{\rm
           tr}}})(1+\cos^2{\gamma_0}+2 \frac{T_{{\rm rot}}}{q T_{{\rm
           tr}}}\cos^2{\gamma_0})}{(1+ \frac{T_{{\rm rot}}}{q T_{{\rm
           tr}}}
     \cos^2{\gamma_0})^2} \right. \\
 \left. + \frac{\eta_0}{2}\left(\frac{\sin{\gamma_0}}{1+\frac{T_{{\rm
             rot}}}{q T_{{\rm tr}}}\cos^2{\gamma_0}}\right. \right.  +
 \left. \left. \frac{\arctan{\left(\sqrt{1 + \frac{T_{{\rm rot}}}{q
               T_{{\rm tr}}}} \cot{\gamma_0}\right)}}{\sqrt{1 +
         \frac{T_{{\rm rot}}}{q T_{{\rm tr}}}} \cos{\gamma_0}} \right)
 \right\}
\end{multline}
\begin{multline}
  \frac{1}{\nu}\frac{d}{dt} T_{{\rm rot}}(t) = + T^{3/2}_{{\rm
      tr}}\frac{\eta_0 }{2 q}\left\{ \frac{2\eta_0(1+\frac{T_{{\rm
            rot}}}{q T_{{\rm tr}}})}{1+\frac{T_{{\rm rot}}}{q
        T_{{\rm tr}}}\cos^2{\gamma_0}} \right.\\
  - \left. \frac{T_{{\rm rot}}}{T_{{\rm tr}}}
    \left(\frac{\sin{\gamma_0}}{1+\frac{T_{{\rm rot}}}{q T_{{\rm
              tr}}}\cos^2{\gamma_0}}\right. \right.  + \left. \left.
      \frac{\arctan{\left(\sqrt{1 + \frac{T_{{\rm rot}}}{q T_{{\rm
                    tr}}}} \cot{\gamma_0}\right)}}{\sqrt{1 +
          \frac{T_{{\rm rot}}}{q T_{{\rm tr}}}}
        \cos{\gamma_0}}\right)\right\} \nonumber
 \end{multline}
 where $\nu=16/3\sqrt{\pi/m}a^2n_0 g(a)$ sets the time scale,
 $\eta_0=\eta\sin{\gamma_0}=[q
 \{1+\beta_0\}/\{2(1+q)\}]\sin{\gamma_0}$, $g(a)$ denotes the pair
 correlation at contact, and $n_0=N/V$.
 
 Rotational energy is conserved only in the perfectly smooth case
 characterized by ($\beta_0 = -1$) or ($\mu = 0$).  Translational
 energy is conserved only if in addition $\epsilon=1$.  The total
 energy is conserved if both translational and rotational energies are
 conserved or in the perfectly rough case ($\mu =\infty \wedge \beta_0
 = +1$) with complete normal restitution $\epsilon = 1$.  For all
 other values of the parameters $\epsilon$, $\beta_0$, and $\mu$ the
 translational and rotational energies decrease with time.

\subsection{Simulations}
Simulations are performed using an event-driven algorithm where the
particles follow an undisturbed translational and rotational motion
until a collision occurs.  In a collision, the particles' velocities
just after contact are computed using the velocities just before
contact as stated in eqs. (\ref{freddy}).  To accelerate the
simulations we use the algorithm of Lubachevsky \cite{Lub91} and a
linked cell structure, which allows us to look for collision partners
in the neighborhood of a given particle only.

To obtain a well-defined initial configuration we start the simulation
on a regular lattice with random velocities chosen from a Boltzmann
distribution and zero angular velocities.  To equilibrate the system
we choose $\epsilon=1$ and $\beta_0=-1$ corresponding to perfectly
{\em smooth} spheres and let the simulation run for 200 collisions per
particle. Then $\epsilon$, $\mu$, and $\beta_0$ are switched to their
desired values.

To circumvent the problem of inelastic collapse, i.e. the time between
two collisions becomes too short to be resolved properly, we use the
$t_{c}$-model \cite{LuNa98}: if the time between a collision and the
preceding one for at least one particle is smaller than a critical
value $t_c$, the collisions parameters are set to their elastic
values.

We are not primarily interested in phenomena like shear and cluster
instabilities, but want to investigate how friction effects the
cooling properties in the rapid flow regime.  Hence we simulate dilute
systems and aim at good statistics.  We perform simulations of 3250
particles with a volume fraction $\rho=\frac{4\pi}{3}a^3\frac{N}{V} =
0.101 $.  For all plots we introduce the dimensionless temperatures
$T=T_{ {\rm tr}}/T_{ {\rm tr}}(0)$ and $R=T_{ {\rm rot}}/T_{ {\rm
    tr}}(0)$ and a dimensionless time $\tau=\nu \sqrt{T_{ {\rm
      tr}}(0)} t~$.  The pair correlation function at contact, $g(a)$
is computed using the Carnahan-Starling formula in 3D \cite{Ha86}:
\begin{equation}
  g(a) = \frac{1-\rho/2}{(1-\rho)^3}~.
\end{equation}
All data presented here corresponds to initially non-rotating
($R(0)=0$, $T(0)=1$) homogeneous spheres ($q=0.4$).

\subsection{Comparison of analytical theory and numerical simulations}

\begin{figure}[htbp]
  \begin{center}
    \leavevmode \epsfig{file=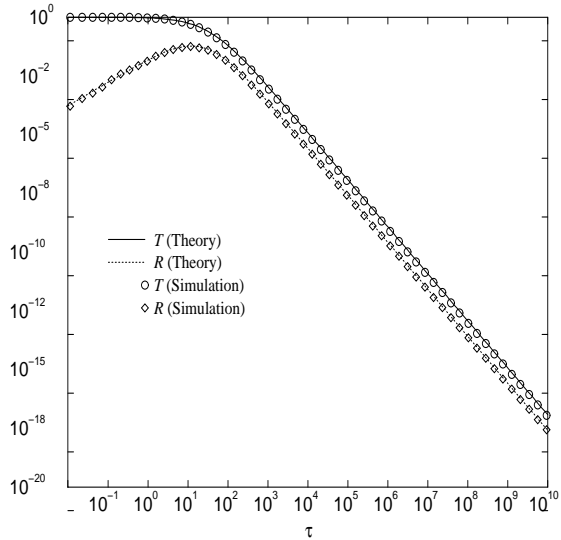,
      width=0.6\textwidth,height=0.6\textwidth}
    \caption{Decay of translational and rotational energy for a set of
      parameter values ($\epsilon = 0.97$, $\mu = 0.092$, $\beta_0 =
      0.44$) corresponding to Soda lime glass; results of the
      approximate analytical theory are compared to data from
      simulations.}
    \label{DataSodaLimeGlass}
  \end{center}
\end{figure}

The most surprising result for the model with impact angle dependent
tangential restitution is a transition which separates two phases with
different asymptotic decays of translational and rotational energies.
For large $\mu$ and nearly complete normal restitution ($\epsilon$
close to 1), we observe cooling properties which are very similar to
those obtained in the model with constant tangential restitution
corresponding to the limit of infinite $\mu$. The asymptotic state is
characterized by a constant ratio of rotational to translational
energy, both decaying according to Haff's law \cite{Haff}.  The time
dependence of the translational and rotational energies for a typical
set of parameters, corresponding to soda lime glass \cite{Foerster},
is shown in Fig.  \ref{DataSodaLimeGlass}.
The numerical solution of eqs. (\ref{dgl1}) is compared to simulations
and good agreement is found for the whole cooling range.  For short
times there is a linear change of both temperatures, whereas in the
asymptotic state both temperatures decay like $t^{-2}$ according to
Haff's law.  In the asymptotic state the ratio of rotational and
translational temperature is constant in time.

\begin{figure}[htbp]
  \begin{center}
    \leavevmode \epsfig{file=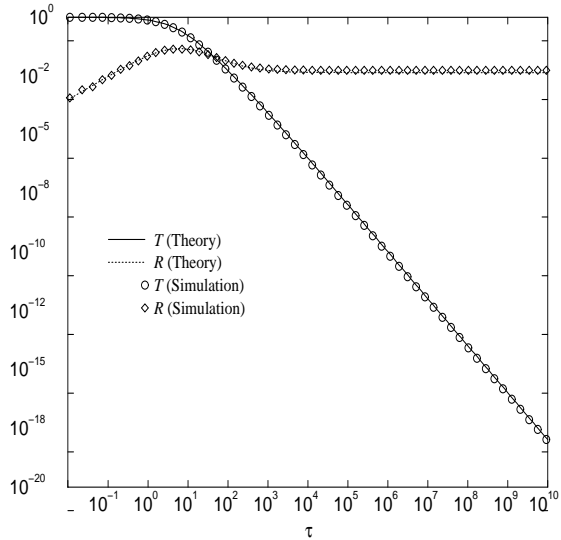,
      width=0.6\textwidth,height=0.6\textwidth}
    \caption{Comparison of analytical theory and simulations for a set
      of parameters $\epsilon = 0.3$, $\mu = 0.2$, and $\beta_0=0.5$
      such that the rotational energy survives and only the
      translational energy decays like $t^{-2}$ (Haff's law).}
    \label{dataEP0:30MU0:20B0:50}
  \end{center}
\end{figure}

For small $\mu$ and small $\epsilon$ the rotational energy remains
constant in time (after an initial increase for small initial $R(0)$).
The translational energy decays according to Haff's law.  An example
is shown in Fig.  \ref{dataEP0:30MU0:20B0:50} for $\epsilon=0.3$ and
$\mu=0.2$.
This, at first surprising result can be explained quite easily:
Coulomb's law of friction yields only very small friction for small
normal loads.  So, when the spheres lose large amounts of their
translational energy (small $\epsilon$) but only a tiny bit of their
rotational energy (small $\mu$) the system develops towards a state in
which hardly any more rotational energy is lost because the grains
only suffer impacts with very small normal load. Hence the {\it
  asymptotic} state resembles that of smooth spheres: The system
consists of a finite fraction of rotating particles at rest. To
discuss this state analytically we expand eqs.  (\ref{dgl1}) for small
$\mu$, which implies $\gamma_0$ close to $\pi$. We set
$\gamma_0=\pi-\delta$ and expand to leading order in $\delta$
\begin{align} 
  \frac{1}{\nu} \frac{d}{dt} T_{{\rm tr}}(t) =& - T^{3/2}_{{\rm tr}}
  ( \frac{1-\epsilon^2}{4} )\\
  \frac{1}{\nu}\frac{d}{dt} T_{{\rm rot}}(t) =&- \frac{\pi \delta
    (1+\beta_0)}{8(1+q)} \frac{T_{{\rm tr}} T_{{\rm rot}}}
  {\sqrt{T_{{\rm tr}}+T_{{\rm rot}}/q}}
\end{align}
In leading order, the translational energy is decoupled from the
rotational energy and equal to the one for smooth spheres. The
solution of the second equation is given by
\begin{equation}
  \sqrt{T_{{\rm rot}}(t)}= const + \frac{8 (1+\beta_0) \pi
    \delta\sqrt{q}}{\nu (1-\epsilon^2)^2 (1+q)}\frac{1}{t}.
\end{equation}
The constant is difficult to evaluate because it depends on the time
scale of the crossover to the asymptotic regime and on the values of
$T_{{\rm tr}}$ and $T_{{\rm rot}}$ on this time scale. Analytical
theory and simulation agree well in this range of parameters, too.
\begin{figure}[htbp]
  \begin{center}
    \leavevmode \epsfig{file=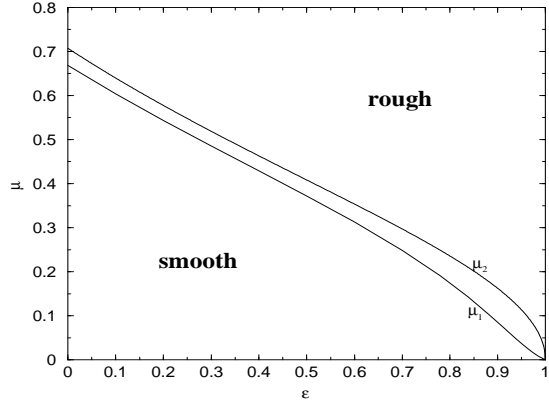,
      width=0.6\textwidth,height=0.45\textwidth}
    \caption{Transition lines $\mu_{1}(\epsilon, \beta_0)$ and
      $\mu_{2}(\epsilon)$ for $\beta_0=0.5$.  While $\mu_{2}$ is
      independent of $\beta_0$, $\mu_{1}$ has a weak dependence on
      $\beta_0$.  The intermediate phase lying between $\mu_{1}$ and
      $\mu_{2}$ is largest for $\beta_0=1$, and $\mu_{1} \to \mu_{2}$
      as $\beta_0 \to 1$.  For $\beta_0=1$ the curve for $\mu_{1}$
      lies less than $0.01$ below the one for $\beta_0=0.5$.}
    \label{CriticalMus}
  \end{center}
\end{figure}
These two regimes with qualitatively different long time behavior are
indicated in Fig. \ref{CriticalMus} as rough and smooth.  Why there
are {\it two} critical values $\mu_{1}$ and $\mu_{2}$ will be
explained in the following discussion.

\begin{figure}[htbp]
  \begin{center}
    \leavevmode \epsfig{file=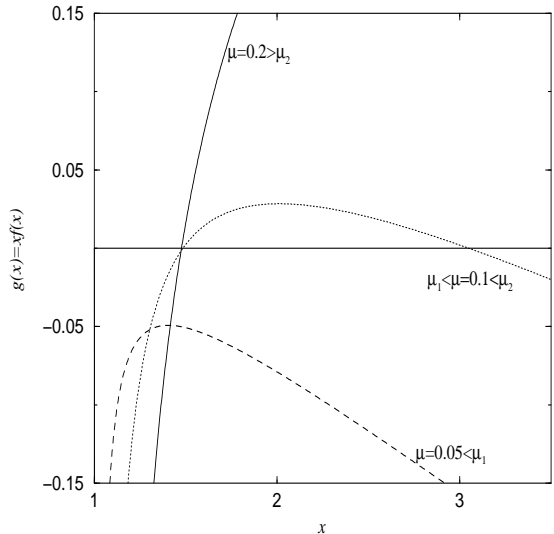,
      width=0.6\textwidth,height=0.6\textwidth}
    \caption{Plot of the function $g(x)$ whose zeros are possible
      asymptotic states for different values of $\mu$.
      ($\epsilon=0.9$, $\beta_0=0.5$).  In this case $\mu_{1}= 0.08507$
      and $\mu_{2}=0.16222$.  For $\mu \gtrless \mu_{2}$,
      $g(x\rightarrow\infty) \rightarrow \pm \infty$.}
    \label{FunktionGvonX}
  \end{center}
\end{figure}

To locate the transition between these two phases, we investigate the
following question: For which range of parameters do eqs. (\ref{dgl1})
allow for a solution with a constant ratio of rotational to
translational energy?  We plug the ansatz $k = T_{{\rm rot}}/T_{{\rm
    tr}} = R/T$ into eqs.  (\ref{dgl1}) and use $k = \frac{d T_{{\rm
      rot}}/dt}{d T_{{\rm tr}}/dt}$.  Introducing $x :=
\sqrt{1+\frac{k}{q}}$ we obtain a function $f(x)$ whose zeros are
possible solutions for an asymptotic state.
\begin{multline}
 \label{asymratio8d}
 f(x)= 2\eta^2 \frac{x^2 }{1+x^2 \cot^2{\gamma_0}}\left(\frac{1 + 2
     x^2 \cot^2{\gamma_0}}{1 + x^2 \cot^2{\gamma_0}} - \frac{1}{q^2
     (x^2-1)}\right) \\+ \eta \frac{1-q}{q} \left(\frac{1}{1+x^2
     \cot^2{\gamma_0}} + \frac{\arctan{(x\cot{\gamma_0})}}{x
     \cot{\gamma_0}}\right) - \frac{1-\epsilon^2}{2}
\end{multline}
In the limit $\mu \rightarrow \infty$ ($\gamma_0 \rightarrow
\frac{\pi}{2}$), the equation $f(x_0)=0$ reduces to a quadratic one
for which exactly one positive zero exists.  We obtain $k= X +
\sqrt{X^2+1}$ where
 \begin{equation} 
 \label{asymratioMARTINx}
 X=\frac{q}{2\eta^2}\left(\frac{1-\epsilon^2}{4} + \eta^2
   \frac{1-q^2}{q^2} -\eta \frac{1-q}{q}\right)
\end{equation}
in agreement with \cite{martin97,Luding,Luding98,Goldshtein}.  For $0
< \mu < \infty$ and $\beta_0 \ne -1$ we get $0 < |\cot{\gamma_0}| <
\infty$.  For $x>1$, consider $f(x)$ in the limits $x \rightarrow 1$
and $x \rightarrow \infty$.
 \begin{equation} 
 \label{limiteins}
 f(x) \underset{x\rightarrow 1}\longrightarrow -\infty
 \end{equation}
 \begin{equation} 
 \label{limitunendlich}
 \lim\limits_{x\rightarrow \infty}{f(x)} =
 \frac{4\eta^2}{\cot^2{\gamma_0}} - \frac{1-\epsilon^2}{2} = (1+\epsilon)^2 \mu^2 - \frac{1-\epsilon^2}{2}
 \end{equation}
 From eqs. (\ref{limiteins}, \ref{limitunendlich}) we see that at
 least one solution $x_0$ for $f(x_0)=0$ and hence for a constant
 asymptotic ratio $R/T=k=q(x^2_0-1)$ exists if
\begin{equation}
  \label{bedingung}
  \mu > \sqrt{\frac{1}{2}\frac{1-\epsilon}{1+\epsilon}} =:
  \mu_{2}(\epsilon)
\end{equation}
This critical value is independent of $\beta_0$.  To find out if there
is more than one solution, and if there are solutions even if eq.
(\ref{bedingung}) is violated we take a look at $g(x):= xf(x)$.  Since
$x>1$, $x_0$ is a zero of $f(x)$ if and only if $x_0$ is a zero of
$g(x)$.  For a given $\beta_0$ there are three qualitatively different
shapes of $g(x)$ depending on $\mu$ (see Fig \ref{FunktionGvonX}).
For small $\mu$ ($\mu<\mu_{1}$), $g(x)$ has no zero.  For
$\mu_{1}<\mu<\mu_{2}$, $g(x)$ has two zeros, and for $\mu>\mu_{2}$,
$g(x)$ has one zero.
When $g(x)$ has two zeros only the {\it smaller} one serves as an
asymptotic ratio, and only if the initial value $R(0)/T(0)$ is smaller
than the {\it greater} zero.  If the initial value is greater than the
{\it greater} zero the system behaves like in the regime where no
solution for an asymptotic ratio exists, that means the rotational
energy survives.

The critical lines $\mu_{1}({\epsilon})$ and $\mu_{2}({\epsilon})$ are
shown in Fig. \ref{CriticalMus} for $\beta_0=0.5$.  $\mu_{2}$ is given
by eq. (\ref{bedingung}) and $\mu_{1}$ is evaluated numerically.  To
conclude, we observe three different phases: 1) a rough phase with a
constant ratio of translational and rotational energies, b) a smooth
phase with constant rotational energy and c) an intermediate phase
where the asymptotic state is determined by the initial value of
$T_{{\rm tr}}(0)/T_{{\rm rot}}(0)$.  The intermediate phase is largest
for $\beta_0=1$, and $\mu_{1} \to \mu_{2}$ as $\beta_0 \to 1$.

\begin{figure}[htbp]
  \begin{center}
    \leavevmode \epsfig{file=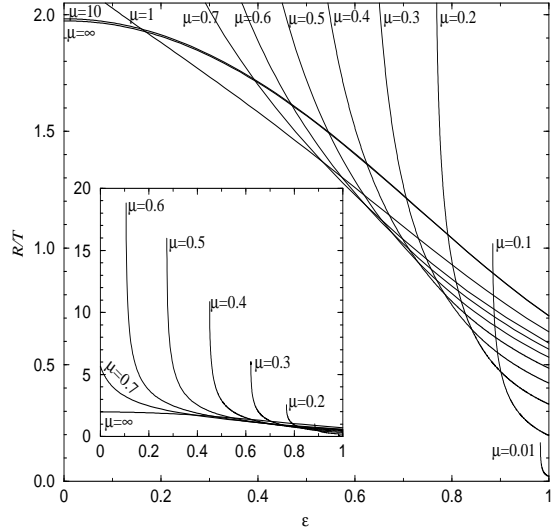,
      width=0.6\textwidth,height=0.6\textwidth}
    \caption{Asymptotic ratio $R/T$ as a function of $\epsilon$ for
      different values of $\mu$, $\beta_0 = 0.5$.  For large $\mu$
      there exists a constant ratio for all $0 \le \epsilon \le 1$.
      For small $\mu$, however, there is a critical $\epsilon$ such
      that for smaller $\epsilon$ the rotational energy survives while
      the translational energy decays like $t^{-2}$.}
    \label{RdTvonEpsilon}
  \end{center}
\end{figure}

For $\beta_0=0.5$ the asymptotic ratio $R/T$ is shown in Fig.
\ref{RdTvonEpsilon} as a function of $\epsilon$ for various $\mu$,
$0.01 \le \mu \le \infty$.  For $\mu>\mu_{1}(\epsilon=0)$ the
asymptotic state is characterized by a constant ratio of rotational
and translational energies for {\it all} values of $\epsilon$.  When
$\mu$ is decreased to smaller values, we find an asymptotically
constant ratio only for sufficiently large $\epsilon$.  We have chosen
$R(0)=0$ so that there is only one transition at $\mu_1$ and we do not
observe the intermediate phase, in which the asymptotic state depends
on the choice of initial condition for $R(0)$.

\begin{figure}[htbp]
  \begin{center}
    \leavevmode \epsfig{file= 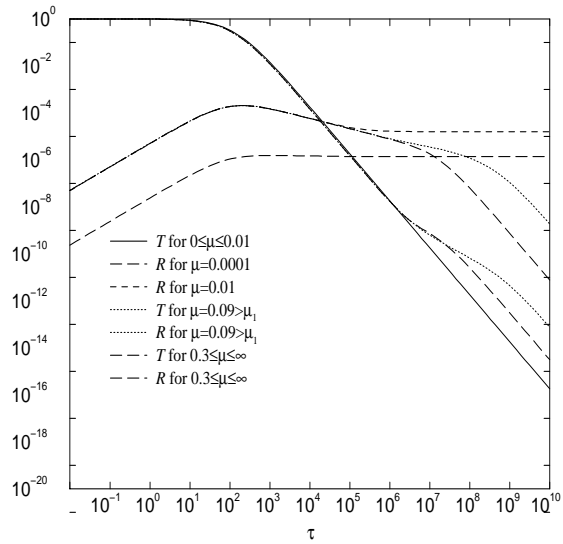,
      width=0.6\textwidth,height=0.6\textwidth}
    \caption{Crossover between the two phases with different long time
      asymptotics as a function of $\mu$ for fixed $\epsilon = 0.97$
      and $\beta_0=-0.99$.  $\mu_{1}= 0.08717$.}
    \label{CrossoverAsFunctOfMu}
  \end{center}
\end{figure}
In Fig. \ref{CrossoverAsFunctOfMu} we show the crossover between the
two phases with different long time asymptotics for the rotational
energy. The coefficients of normal and tangential restitution are
fixed ($\epsilon = 0.97$, $\beta_0=-0.99$) while $\mu$ is varied over
the full range $0<\mu<\infty$. For small $\mu$ the translational
energy is almost the same as that for smooth spheres and the
rotational energy is constant for long times. For $\mu>\mu_{1}$ the
rotational energy decays, the sooner the larger $\mu$. The decay of
the translational energy is slowest for $\mu\to\mu_{1}^+$. As $\mu \to
\infty$ it approaches the curve for constant tangential restitution.

\begin{figure}[htbp]
  \begin{center}
    \leavevmode
    \epsfig{file=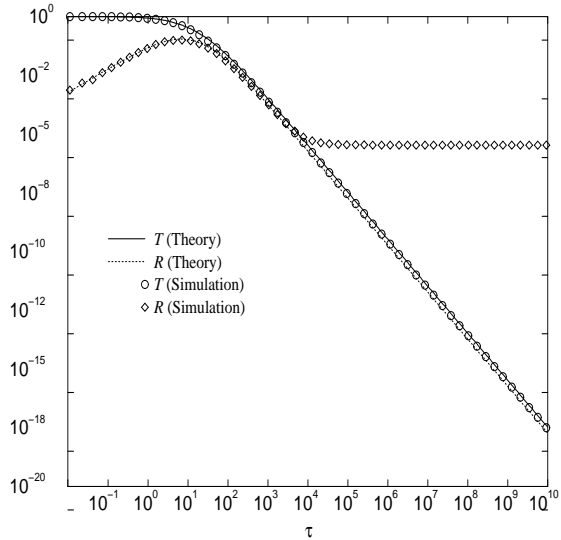, width=0.6\textwidth,height=0.6\textwidth}
    \caption{Intermediate regime of parameters ($\epsilon = 0.87$,
      $\mu = 0.25$, $\beta_0 = 0.43$) corresponding to cellulose
      acetate; theory predicts a constant ratio for translational and
      rotational energies, wheras simulations show a time persistent
      rotational energy.}
    \label{dataCelluloseAcetate}
  \end{center}
\end{figure}

\begin{figure}[htbp]
  \begin{center}
    \leavevmode \epsfig{file=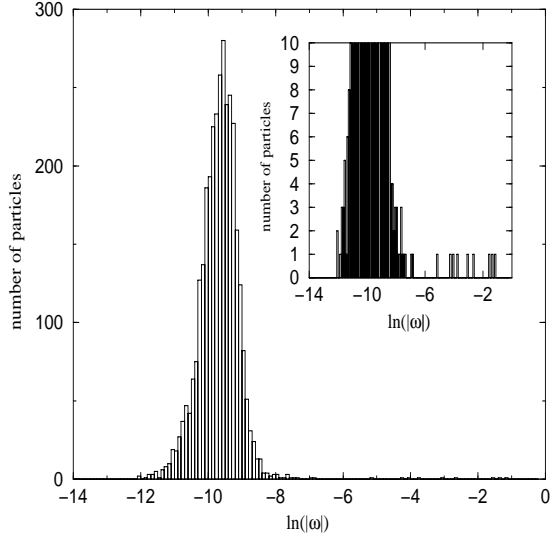,
      width=0.6\textwidth,height=0.6\textwidth}
    \caption{Histogram at $\tau = 10^{6}$ of
      $\ln{(|\boldsymbol{\omega}|)}$ for cellulose acetate ($\epsilon
      = 0.87$, $\mu = 0.25$, $\beta_0 = 0.43$)}
    \label{Histogramm}
  \end{center}
\end{figure}

Deviations between the approximate analytical theory and simulations
are observed in the parameter regime, close to the transition lines.
In particular, the parameters $\mu$ and $\epsilon$ can be chosen such
that the analytical theory predicts a Haff type decay of the
rotational energy, whereas the simulations reveal constant $R$.  In
Fig. \ref{dataCelluloseAcetate} we show results of a simulation for
the parameters of cellulose acetate spheres as measured by Foerster et
al.  \cite{Foerster}. Looking at single grains in the simulation, one
finds extremely non-Gaussian states, in the sense that few particles
rotate with high angular velocities and dominate the average
rotational energy.  In Fig. \ref{Histogramm} we plot a histogram of
the rotational velocities for a snapshot taken at $\tau = 10^6$.  It
reveals clearly that the rotational energy is dominated by few
particles with high rotational velocities.  Snapshots taken at other
instants of time show that the identity of particles with high
rotational velocities is conserved. The HCS approximation is not
expected to hold in such a state which is strongly non-Gaussian and
has a nearly log normal distribution of the angular velocity.

\section{Modification of Coulomb's law for low load}

The persistence of rotational energy during cooling can be traced back
to Coulomb's law which predicts vanishing frictional losses for
grazing collisions, regardles of the magnitude of the relative
tangential velocity of the contact point (see Fig.
\ref{BetaVonGamma}).  For realistic materials one would expect some
residual friction due to surface roughness. This effect can be modeled
crudely by a minimal roughness $\beta_{\rm min}$ such that
$\beta(\gamma) \ge \beta_{\rm min}$.  In addition, Coulomb's law,
$|\boldsymbol{F}_{\rm fric}| = \mu
|\hat{\boldsymbol{n}}\cdot\boldsymbol{F}_{\rm load}|$, which we have
used only holds for sufficiently large normal loads.  When the normal
load gets very small, as it happens in the late stages of cooling and
in particular for small $\epsilon$, cohesion begins to play a role so
that there will always be friction even for zero-load impacts as
discussed by Johnson, Kendall, and Roberts \cite{Johnson1971}. They
modify Coulomb's law according to:
 \begin{equation}\label{JKR}
 |\boldsymbol{F}_{\rm fric}| = \mu
 \left(|\hat{\boldsymbol{n}}\cdot\boldsymbol{F}_{\rm load}| + F_0 +
   \sqrt{2 |\hat{\boldsymbol{n}}\cdot \boldsymbol{F}_{\rm load}| +
 F_0^2}\right)
 \end{equation}
 where $F_0 = \frac{3}{2} \pi a E_{\rm s}>0$ and $E_{\rm s}$ denotes
 the material-specific surface energy.  Even simpler is the following
 version discussed as early as 1934 \cite{old}
 \begin{equation}\label{coul1938}
 |\boldsymbol{F}_{\rm fric}| = \mu
 \bigg(|\hat{\boldsymbol{n}}\cdot\boldsymbol{F}_{\rm load}| + F_{0}
 \bigg) 
 \end{equation}
 with a small positive quantity ($F_0$) due to cohesion.  To estimate
 the effects of cohesion we integrate eq. (\ref{coul1938}) over the
 duration of a collision $\Delta t \sim a c^{-4/5}v^{-1/5}$ as given
 by the theory of Hertz \cite{Hertz} ($c$ denotes the velocity of
 sound).  We then obtain a modified Coulomb law
\begin{equation}
m|\hat{\boldsymbol{n}} \times \boldsymbol{\Delta v}| =\mu(
m|\hat{\boldsymbol{n}} \cdot \boldsymbol{\Delta v}| + F_0 \Delta t)
\end{equation} 
which corresponds to an impact angle dependent coefficient of
tangential restitution for sliding contacts
\begin{equation}
  \beta^{*}(\gamma) = -1 + \frac{1+q}{q} \mu \left\{
    (1+\epsilon)|\cot{\gamma}| +\frac{2F_0 \Delta t}{m|\boldsymbol{g}
      \times \hat{\boldsymbol{n}}|}\right\}.
 \end{equation}
 and generalizes eq. (\ref{betagamma}) to include cohesion forces for
 low impact collisions. The most important feature is a finite
 coefficient of tangential restitution for impact angles
 $\gamma=\pi/2$.
 
\begin{figure}[htbp]
  \begin{center}
    \leavevmode
    \epsfig{file=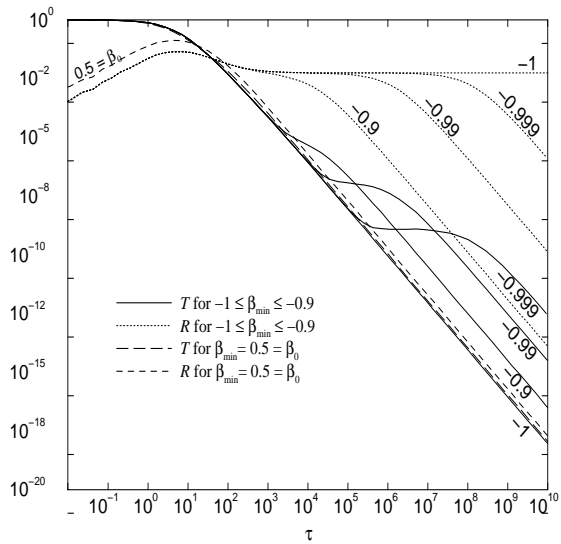, width=0.6\textwidth,height=0.6\textwidth}
    \caption{Simulations for $\epsilon = 0.3$, $\mu = 0.2$, and
      $\beta_0=0.5$ for different $\beta_{\rm min}$.  The values of
      $\beta_{\rm min}$ are shown at the corresponding curves.  Any
      $-1<\beta_{\rm min}\le\beta_0$ causes the rotational energy to
      decay. As $\beta_{\rm min}$ approaches -1 the plateau of the
      rotational energy persists for longer times.  $\beta_{\rm
        min}=\beta_0$ cancels all $\mu$-dependence of $\beta$ and thus
      reveals a constant $\beta$.  }
    \label{DifferentBetamins}
  \end{center}
\end{figure}

 The question arises, whether the time persistence of the rotational
 energy survives, if finite $\beta(\frac{\pi}{2})$ is taken into
 account as predicted by surface roughness as well as by cohesion
 forces. In Fig. \ref{DifferentBetamins} we show results of
 simulations with an angle dependent coefficient of restitution, as
 given in eq.  (\ref{betax}), however, with a lower bound $\beta_{\rm
   min}$ as predicted by surface roughness as well as cohesion forces.
 The rotational energy shows a plateau for intermediate times and
 decays asymptotically like $t^{-2}$ for long times.  The length of
 the plateau and the onset of the decay depend on the value of
 $\beta_{\rm min}$ as expected: The plateau is longer and the decay
 sets in at later times the closer $\beta_{\rm min}$ is to $-1$.

\section{Summary and Outlook}

We have investigated the effects of friction on the cooling properties
of granular particles.  We observe three distinct phases which differ
qualitatively in their late stage of cooling. In the {\it rough} phase
cooling is characterized by a constant ratio of translational and
rotational energies whereas the {\it smooth} phase is characterized by
a time persistent rotational energy even for the latest times.  These
two regimes are separated by an intermediate regime in which the late
stage of cooling can be either smooth or rough depending on the
initial conditions.  Both regimes are also observed in the
simulations. In fact, approximate analytical theory and simulation
agree well within both phases.  Close to the intermediate regime we
find a strongly non-Gaussian angular velocity distribution which
causes the analytical theory to fail.  Deviations between theory and
simultaion may also be due to finite size effects which are expected
to also play a role in experiments on granular media, in contrast to
experiments on conventional systems of statistical mechanics, where
finite size effects are usually negligible.

Any small friction for grazing collisions as generated for example by
surface roughness causes the rotational energy to decay. However, a
plateau survives for intermediate time scales, such that the decay of
the rotational energy sets in at much later times than the decay of
the translational temperature.  The length of the plateau of the
rotational energy, and hence the time of decay diverges as the
roughness goes to zero.

A possible extension of our work are driven systems. A particular way
of driving - adding random velocity vectors - has been investigated
recently with simulations and approximate analytical theory by Cafiero
et al. \cite{Cafiero}. They find an asymptotic state in which both
kinetic energies take on constant values due to driving.

\section*{Acknowledgements}
We thank Stefan Luding for useful discussions.  This work has been supported by the DFG through SFB 345 and Grant
Zi$209/6-1$.

\end{document}